\documentclass[aps,prd,reprint, superscriptaddress, nofootinbib]{revtex4-2}

\usepackage{graphicx, subfigure}        % Including figure files
\usepackage{aas_macros} 
\usepackage{amssymb,amsmath,placeins,epsfig,array,bm}
\usepackage[usenames,dvipsnames]{color}
\usepackage[normalem]{ulem}
\usepackage[breaklinks,colorlinks,urlcolor=blue,citecolor=blue,linkcolor=blue]{hyperref}
\usepackage{float}
\usepackage[T1]{fontenc}
\usepackage{times}
\usepackage{soul}
\newcommand{\IVEi}{IVE$_{\mathrm{virial}}$}
\newcommand{\IVEo}{IVE$_{\mathrm{infall}}$}
\newcommand{\rvir}{$r_\mathrm{200m}$}
\newcommand{\bol}[1]{\boldsymbol{#1}}

\begin{document}

\title{Discovering the building blocks of dark matter halo density profiles with neural networks}

\author{Luisa Lucie-Smith}
\email[]{luisals@mpa-garching.mpg.de}
\affiliation{Max-Planck-Institut für Astrophysik, Karl-Schwarzschild-Str. 1, 85748 Garching, Germany}

\author{Hiranya V. Peiris}
\affiliation{Department of Physics \& Astronomy, University College London, Gower Street, London WC1E 6BT, UK}
\affiliation{The Oskar Klein center for Cosmoparticle Physics, Stockholm University, AlbaNova, Stockholm, SE-106 91, Sweden}

\author{Andrew Pontzen}
\affiliation{Department of Physics \& Astronomy, University College London, Gower Street, London WC1E 6BT, UK}

\author{Brian Nord}
\affiliation{Fermi National Accelerator Laboratory, P.O. Box 500, Batavia, IL 60510, USA}
\affiliation{Department of Astronomy and Astrophysics, University of Chicago, Chicago, IL 60637, USA}
\affiliation{Kavli Institute for Cosmological Physics, University of Chicago, Chicago, IL 60637, USA}

\author{Jeyan Thiyagalingam}
\affiliation{Scientific Computing Department, Rutherford Appleton Laboratory, Science and Technology Facilities Council, Harwell Campus, Didcot, OX11 0QX}

\author{Davide Piras}
\affiliation{Department of Physics \& Astronomy, University College London, Gower Street, London WC1E 6BT, UK}

\date{\today}

\begin{abstract}
The density profiles of dark matter halos are typically modeled using empirical formulae fitted to the density profiles of relaxed halo populations. We present a neural network model that is trained to learn the mapping from the raw density field containing each halo to the dark matter density profile. We show that the model recovers the widely-used Navarro-Frenk-White (NFW) profile out to the virial radius, and can additionally describe the variability in the outer profile of the halos. The neural network architecture consists of a supervised encoder-decoder framework, which first compresses the density inputs into a low-dimensional latent representation, and then outputs $\rho(r)$ for any desired value of radius $r$. The latent representation contains all the information used by the model to predict the density profiles. This allows us to interpret the latent representation by quantifying the mutual information between the representation and the halos' ground-truth density profiles. A two-dimensional representation is sufficient to accurately model the density profiles up to the virial radius; however, a three-dimensional representation is required to describe the outer profiles beyond the virial radius. The additional dimension in the representation contains information about the infalling material in the outer profiles of dark matter halos, thus discovering the splashback boundary of halos without prior knowledge of the halos' dynamical history.
\end{abstract}

%\keywords{}

\maketitle

\section{Introduction}
In the standard cosmological model, dark matter accumulates in stable, virialized `halos', which form the building blocks of cosmic large-scale structure and wherein galaxy formation takes place. The density structure of dark matter halos contains key information about cosmology and the nature of dark matter \cite{Frenk1985, Frenk1988, Dubinski1991}. High-resolution $N$-body simulations reveal that the spherically-averaged density of these halos declines with radius from $\rho \propto r^{-1}$ in the inner regions to $\rho \propto r^{-3}$ in the outskirts \cite{Navarro1996, Navarro1997}. This functional form provides a good fit to halos over two decades in radius for a large range of halo masses and for several different cosmological models \cite{Wang2020}. Density profiles with similar forms have been shown to arise even in the absence of hierarchical growth, for example from hot dark matter initial conditions or even from spherical collapse \cite{Huss1999, Wang2009, Moore1999, AvilaReese1998}. This suggests that universal density profiles are a generic feature of gravitational collapse.

The physical origin of this near-universal shape is still not well understood. Many attempts at providing an explanation from first principles have been put forward, for example invoking the role of mergers \cite{Salvador1999, Salvador2012},  adiabatic invariants \cite{Dalal2010}, or the generation of entropy \cite{Pontzen2013} as being responsible for halo structural similarity. The lack of a consensus on the origin of self-similar density profiles means that the modeling of profiles relies on empirically-found fitting formulae \cite{Navarro1996, Navarro1997, Einasto1965}. These are tested on the density profiles of selected, dynamically-relaxed halo populations. Theoretical efforts have focused on the connection between the free parameters of the fitting functions and halo properties such as their formation time, their mass accretion histories, as well as the fitting function parameters' dependence on cosmology \cite{Bullock2001, Correa2015, Brown2021}.

We present a novel approach for learning the mapping from the raw density field containing each halo to the dark matter density profile. The goal is to recover the independent set of components needed to describe the profiles. Here, we investigate whether neural networks can be used to discover such components from unprocessed raw data. We design a supervised encoder-decoder architecture that compresses the information in the 3D density field containing dark matter halos into a compact low-dimensional latent representation. The latent representation and any given value of radius $r$ are then mapped to the spherically-averaged density $\rho(r)$. The compression step and the subsequent prediction step are both implemented by neural networks. This architecture allows us to extract knowledge about the underlying physics from the neural network: the representation contains all the information used by the neural network to predict the density at any radius $r$ of the profile. Therefore, interpreting the representation reveals what components are required for modeling the density profiles of dark matter halos. Our network architecture was inspired by \textit{SciNet} \cite{Iten2020}, a neural network able to rediscover the known parameters in various 1D toy examples. Here, we consider a complex real-world scenario which deals with 3D inputs and where the physically relevant quantities are not known a priori. 

A key aspect of our work is the ability to interpret the latent representation discovered by the neural network. To do so, we require that each latent component captures different, independent factors of variation in the profiles. This requirement reflects the idea that physically relevant parameters describe aspects of the system that can be varied independently. This property is formally known as \textit{disentanglement} in the context of representation learning \cite{Bengio2013}. Discovering disentangled representations of the data has attracted attention, as they offer a number of advantages, including interpretability \cite{Higgins2017, kumar2017, Kim2018, eastwood2018, Chen2018}. However, there is currently no general consensus on the correct measure of disentanglement: different proposed metrics impose different implicit assumptions which may not generalize to complex real-world scenarios \cite{Zaidi2020, Sepliarskaia2021}. In this work, we use the information-theoretic metric of \textit{mutual information} between the different latent components to assess the degree of disentanglement. Mutual information has been used to measure the entanglement between the latent parameters and the known ground-truth latent factors \cite{Chen2018}, or between the latent variables and the observed data in the context of generative adversarial networks (GANs) \cite{Chen2016}.

Our machine-learning framework is designed to predict halo profiles starting from the full information about the inner structure of halos given by the 3D density field around the center of the halo. This presents a generalization over existing analytic fitting formulae, which are instead fitted on preprocessed information about the halos' inner structure i.e., the spherically-averaged density profiles themselves. Thus, our machine-learning framework is not limited to spherically-averaged densities, but can be adapted to predict other halo observables. In this work, we focus on spherically-averaged density profiles as a first application, which allows us to test our model against widely-used existing analytic profiles.

In a variety of problems in physics, interpretable machine-learning frameworks have been used to discover relations within the data that can be interpreted with respect to the underlying physics of the system \cite{LucieSmith2018, LucieSmith2019, Iten2020, LucieSmith2020, DAgnolo2019}. Most of these methods require prior knowledge of the system of interest, for example a priori knowledge of the relevant variables or the underlying dimensionality. Recently, \citet{Sedaghat2021} adopted a similar architecture to \textit{SciNet} in an unsupervised setting, where a neural network is trained to find a low-dimensional representation of stellar spectra. Similar to our work, they used mutual information for interpretability; however, their use of mutual information was limited to identifying potential correlations between the latent representation and previously known parameters. They found that two of the latent components discovered by the neural network resemble known stellar physics parameters, but do not provide interpretations of the four remaining relevant latent components. Our work takes mutual information one step forward in the context of interpretability: we not only use it to formalize the concept of disentanglement, but also to interpret the discovered latent representation without the need to compare to known parameters. We additionally compute the mutual information between the latent representation and the parameters of existing density profile fitting formulae for comparison.

We present an overview of the neural network framework in Sec.~\ref{sec:overview}, and provide details on the trained data and the neural network model in Sec.~\ref{sec:data} \& \ref{sec:model}, respectively. We show the predictive performance of our trained model in Sec.~\ref{sec:predictions} and then move to interpreting the latent representation in Sec.~\ref{sec:latents}. We draw our final conclusions in Sec.~\ref{sec:conclusions}.

\section{Overview of the model}
\label{sec:overview}
\begin{figure*}
	\centering
	\includegraphics[width=0.95\textwidth]{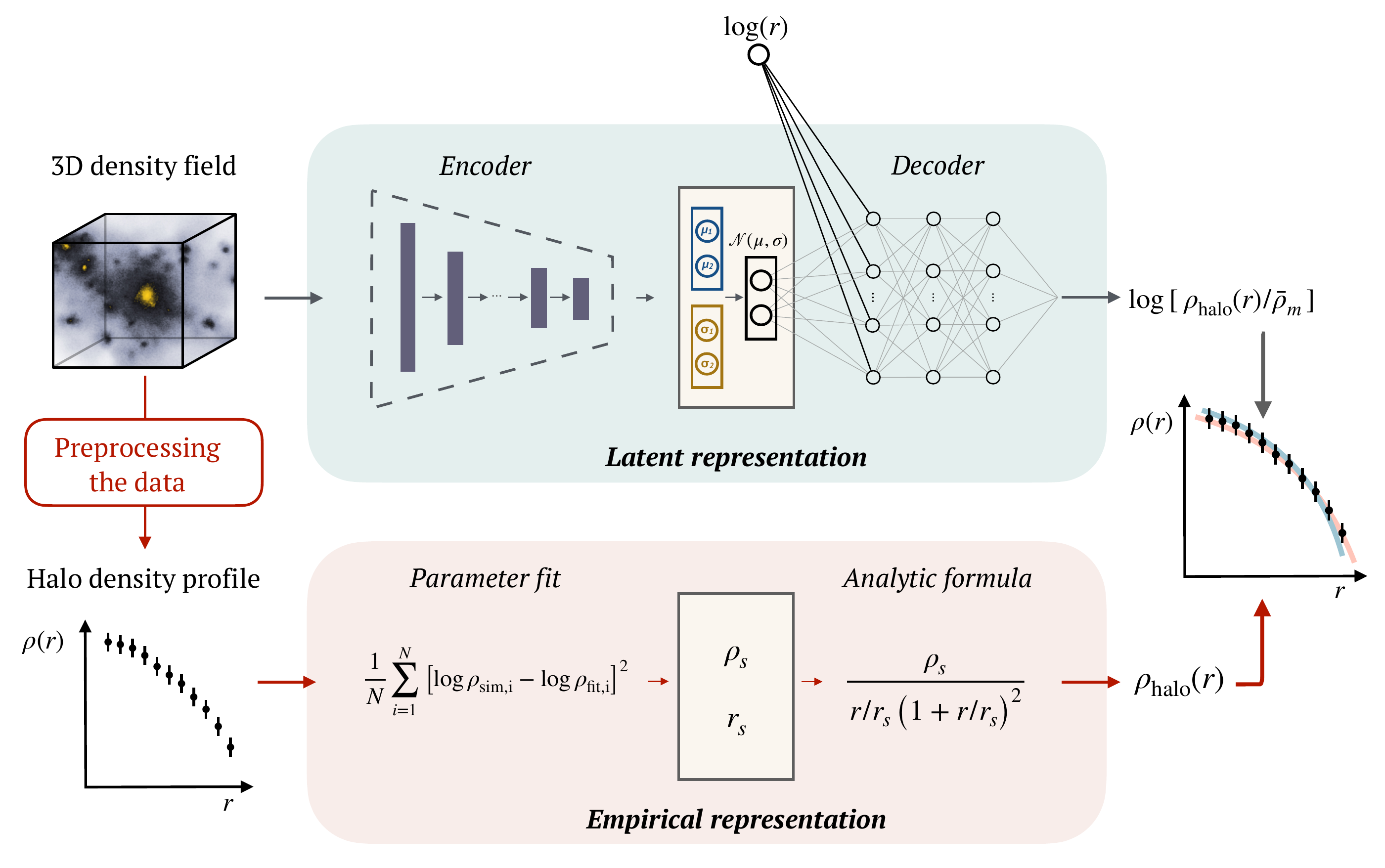}
	\caption{The interpretable variational encoder (IVE) consists of an encoder compressing the 3D density field containing each halo into a low-dimensional latent representation, followed by a decoder mapping the latent representation and a given value of $r$ to the spherically-averaged density $\rho(r)$. In this illustration, the latent space is two-dimensional; however, the dimensionality of the latent space can be increased to any arbitrary value. The latent representation only retains the information required by the model to predict the halo density profiles, allowing us to interpret the representation as independent factors of variability in the density profiles. The decoder plays a role similar to an analytic fitting formula, which takes as input a set of halo-specific parameters and returns $\rho(r)$ for any given $r$. The encoder is equivalent to the parameter fitting procedure, in that they return those halo-specific parameters used by the analytic formula (decoder). However, the inputs to the encoder and the fitting procedure are fundamentally different: the former extracts information directly from the 3D density field containing the halo, whereas the latter uses processed information of that field, i.e. the spherically-averaged density profiles themselves. The latter data processing is a step motivated by human intuition and physical frameworks such as the secondary infall model \cite{GunnGott1972}.}
	\label{fig:architecture}
\end{figure*} 

We adopt an encoder-decoder framework designed for the purpose of knowledge extraction: we wish to gain new insights from a deep learning model by extracting information about the underlying physics of the problem of interest from its latent parameters and its outputs. A schematic illustration of our framework is shown in the top panel of Fig.~\ref{fig:architecture}. 

The encoder compresses the 3D input into a lower-dimensional \textit{latent representation} via a 3D deep convolutional neural network (CNN). The representation consists of an $L$-dimensional Gaussian distribution; the encoder returns the means and variances of each latent component. This means that each halo has a range of possible representations described by the multivariate Gaussian latent distribution. The \textit{decoder} then maps a single realization of the representation and an additional input given by the log-radius $\log(r)$ to the output $\rho(r)$. We denote the additional input to the decoder --  the log-radius $\log(r)$ -- as the \textit{query}. This query was chosen because it is the argument of the function we aim to learn, $\rho(r)$. Our encoder-decoder architecture closely resembles that of variational autoencoders, except that our setting is supervised, whereas standard variational autoencoders are unsupervised. We refer to our model as an \textit{interpretable variational encoder} (IVE), since it adopts elements of variational autoencoders, while its supervised nature allows for interpretability. The IVE resembles the design of \textit{SciNet}, but with enhanced elements of interpretability provided by the mutual information measure, allowing the framework to generalize to scenarios where the physically relevant quantities are not known a priori.

Unsupervised variational autoencoders are encoder-decoder models trained to first compress the input data into a latent representation, and then reconstruct the input data from the representation. The latent representation must therefore capture all the information required to reconstruct the input data completely. Here, our goal is not to reconstruct the entire input data but rather a processed version of it. Therefore, unlike in unsupervised variational autoencoders, the latent representation need not describe the input data completely; it only needs to capture the information necessary to predict the spherically-averaged density profile. This is achieved through the query: it induces the latent space to retain the information used by the IVE to predict $\rho(r)$ for any given value of $r$, by construction. The presence of the $\log(r)$ query therefore plays a crucial role in enabling interpretability. Without it, the information about the density profile would be spread throughout the parameters of the whole encoder, thus losing the ability to interpret the model. 

For the IVE to produce interpretable representations, we further constrain it to satisfy certain desired properties. We require that the latent components be statistically independent from each other, reflecting the idea that physically relevant parameters describe independent factors of variation in a system. Under this independence assumption, the network is then encouraged to choose a representation that stores different relevant factors in different latents. We implement these requirements in the loss function (see Sec.~\ref{sec:model}), as done when disentangling standard variational autoencoders \cite{Higgins2017, Burgess2018}.

Figure~\ref{fig:architecture} compares the IVE model with standard analytic fitting formulae such as the Navarro–Frenk–White (NFW) profile \cite{Navarro1997}. The decoder of the IVE plays a similar role to the analytic NFW formula: these both take as input a set of halo-specific parameters and return the $\rho(r)$ for any given $r$. The encoder is the equivalent of a $\chi^2$-fitting procedure, in that both return the best-fit parameters for a given halo. A fundamental difference between the two methods lies in their respective inputs: the IVE is provided with the raw  3D density field containing the halo, whereas the analytic profile is fitted on preprocessed information about the halo structure i.e., the spherically-averaged density profiles themselves. Therefore, our work is not limited to spherically-averaged densities but can be broadly applied to other halo properties.

In summary, the IVE consists of an encoder mapping the 3D density field containing each halo to a latent representation, followed by a decoder mapping the latent representation and a given value of $r$ to the spherically-averaged density $\rho(r)$. Our architecture is both \textit{interpretable} and \textit{explainable}, which we define as follows. Interpretability concerns the ability to produce outputs that help us understand the inner workings of machine-learning models and how the models reach their final predictions; this is achieved using the information-theoretic metric of \textit{mutual information}. Explainability denotes the ability to map the interpretations onto existing knowledge in the relevant science domain; this is achieved by evaluating the mutual information between the latent variables and the halos' density profiles. This directly reveals the information content of each latent variable in relation to the halos' density profiles (see Sec.~\ref{sec:latents}). 

\subsection{Probing different radial ranges with two IVEs}
Our goal is to use neural networks to discover the independent degrees of freedom in dark matter density profiles over a wide range of radii. In particular, we wish to probe the typical radial range covered by existing analytic fitting formulae, as well as regions out to larger radii where the halo joins into the surrounding large-scale structure. Existing analytic formulae are typically only fitted to the density profiles up to the halo boundary, for example defined as the radius which encloses a mean density that is 200 times the mean matter density of the Universe, \rvir{}. On the other hand, halo outskirts have gained recent interest as they are sensitive to the halos' mass accretion rate, as well as being sensitive to the nature of dark matter, dark energy and modified gravity theories \cite{Banerjee2020, Adhikari2018}.

Whether or not the IVE model can accurately predict the density profile over the desired range of radii crucially depends on the scales probed by the 3D input data. In other words, the IVE must have access to the 3D density field at scale $r$ in order to infer the spherically-averaged density $\rho(r)$. The inputs are given by the density field within a cubic sub-box of the simulation centered on the halo; the scales that are accessible to the IVE are therefore set by the volume and resolution of the sub-box. 

Due to memory limitations of current state-of-the-art GPUs, we were unable to generate input sub-boxes with high-enough resolution and large-enough volume for the IVE to fit density profiles over the entire radial range. To overcome this technical limitation, we trained two independent IVE models: the first one is used to model the density profile up to the halo boundary ($r_\mathrm{max} =$ \rvir{}) and the second to model profiles beyond the halo boundary ($r_\mathrm{max} =2$\rvir{}). We refer to the first model as \IVEi{} and the second as \IVEo{}. The \IVEi{} model provides a benchmark to compare to existing analytic models, such as the NFW and Einasto profiles, that are also valid only within the boundary of the halo. The \IVEo{} model instead is used to uncover features of the less explored outer profile of halos.

\section{The simulated data}
\label{sec:data}
\subsection{The simulations}
We generated the training data from four dark matter-only $N$-body simulations produced with GADGET-4 \citep{gadget4}, each consisting of a box of size $L = 50 \, \mathrm{Mpc} \, h^{-1}$ and $N=512^3$ simulation particles evolving from $z = 99$ to $z = 0$. We made use of \textsc{pynbody} \citep{pynbody} to analyze the information contained in the simulation snapshots. The simulations adopt a \textit{Planck} $\Lambda$CDM cosmological model \citep{Aghanim2020}. Each simulation is based on a different realization of a Gaussian random field drawn from the initial power spectrum of density fluctuations, generated using \textsc{genetIC} \citep{Stopyra2021}. The softening length is $\epsilon = 1 \, \mathrm{kpc} \, h^{-1}$.

Dark matter halos were identified at $z=0$ using the SUBFIND halo finder \citep{gadget2, gadget4}, a friends-of-friends method with a linking length of 0.2, with the additional requirement that particles in a halo be gravitationally bound. We restricted our analysis to halos within the mass range $\log ( M/M_{\odot} ) \in [11, 13]$, in order to fully resolve the inner profile of the lowest-mass halos and not be affected by small-number statistics at the high-mass end.

\subsection{Density profile outputs}

We used the $z=0$ snapshot of the simulations to assign to each halo its ground-truth density profile. We used the halo finder to identify the centers of the halos in the simulation. For every halo, we computed its density profile by evaluating the density within 24 bins in radius, logarithmically-spaced in the range $r \in [3\, \epsilon, 2\, r_\mathrm{200m}]$, where $\epsilon$ is the softening length. The density $\rho(r)$ is computed using all particles at distance $r$ from the halo center, not just those belonging to the halo according to the halo finder. The lower radial bound, $r_\mathrm{min} = 3\, \epsilon$, is the smallest scale one can trust before reaching scales that are affected by the gravitational softening of the simulation. The upper radial bound, $r_\mathrm{max} = 2\, r_\mathrm{200m}$ was chosen in order to probe the outer profile of the halo beyond the virial radius. From the density profiles, we then assigned to each halo a set of query-ground truth pairs as follows. The queries are given by the centers of every $i$-th radial bin, $\log (r_i )$; the ground-truth labels are given by $\log [\rho(r_i)/\bar{\rho}_m]$, where $\rho(r_i)$ is the density evaluated at query $r_i$ and $\bar{\rho}_m$ is the mean matter density of the Universe. 

In contrast to $r_\mathrm{min}$ which is fixed for all halos in the simulation, the upper bound set by $r_\mathrm{200m}$ naturally varies for every halo. This implies that the set of query values changes for every halo, meaning that the algorithm must also learn about which physical scales are relevant for any given halo.

The outputs of the \IVEi{} and \IVEo{} models differ by the choice of radial bins used for training. For the \IVEi{} model, we restricted the outputs to the first 21 bins up to $r_\mathrm{max} = r_\mathrm{200m}$. For the \IVEo{} model, we used all bins up to $r_\mathrm{max} = 2\, r_\mathrm{200m}$, except for the innermost one.

\subsection{3D density field inputs}
\label{sec:inputs}

The inputs are generated from the 3D density field, $\rho(\mathbf{x})$, at $z=0$. For each halo, the input is given by $\log [\rho(\mathbf{x})/\bar{\rho}_m + 1]$ in a cubic sub-region of the full simulation of size  $L_\mathrm{sub-box}$ and resolution $N_\mathrm{sub-box}$, centered on the halo center. All halos, independently of their size and mass, have input sub-boxes of the same size and resolution. The density field was constructed from the position of particles in the simulation using a smoothed-particle hydrodynamics (SPH) procedure, and evaluated at each voxel of the cubic sub-box. 

The inputs of the \IVEi{} and \IVEo{} models differ by the size and resolution of the sub-boxes. For the \IVEi{} model, we chose a $N=131^3$ sub-box of size $L_\mathrm{sub-box} = 0.4 \, \mathrm{Mpc} \, / \, h$. This choice ensures that the inputs have access to the relevant scales: the voxel size, $l \sim 3\, \mathrm{kpc} \, / \, h$, matches the smallest radial value of the profile, and the sub-box size is $2\times$ larger than the virial radius of $87\%$ of halos. The remaining $13\%$ of halos have a larger virial radius up to \rvir{}$\sim L_\mathrm{sub-box}$. However, the density on these large scales is highly correlated with that at $r = r_\mathrm{200m}$; we therefore expect the algorithm to be able to make sensible predictions up to the virial scale of the largest halos even if its largest accessible scale is lower than the latter. For the \IVEo{} model, we chose a $N=131^3$ sub-box of size $L_\mathrm{sub-box} = 0.6 \, \mathrm{Mpc} \, / \, h$. In this case, we restricted our halo population to halos with \rvir $\leq 150\, \mathrm{kpc} \, / \, h$ so that the input sub-box is $2\times$ larger than the largest scale of interest in the profile, i.e. $r_\mathrm{max}=2\,$\rvir{}, for all halos.

\section{The Interpretable Variational Encoder}
\label{sec:model}
The IVE architecture has two main components: the encoder, mapping the 3D input sub-boxes to a latent representation, and the decoder, mapping the latent representation and the query $r$ to the output $\rho(r)$. 

The encoder is a 3D CNN that consists of a series of convolutional layers, in which the algorithm learns to extract relevant features from the input data. Feature extraction in CNNs is hierarchical: the first layers learn local, low-level features, which are then combined by subsequent layers into more global, higher-level features. Features are extracted by performing convolutions between the input and a number of kernels in every layer, such that each kernel learns to detect a specific type of feature present in the input. We used five convolutional layers, with 16, 16, 32, 32, 32 kernels for the five convolutional layers, respectively, all of size $3\times3\times3$. All convolutional layers are followed by a non-linear leaky rectified linear unit (Leaky ReLU) \citep{Nair2010} activation function and a subsequent max-pooling layer. The pooling layer decreases the resolution of the 3D outputs of the convolutional layer, by taking the maximum value in small ($2\times2\times2$) regions. In summary, the encoder of the IVE consists of a model with parameters $\phi$ (weights and biases) mapping the inputs $\bol{x}$ to a multivariate distribution in the latent space $p_{\phi}(\bol{z} | \bol{x})$. We assume that it is possible to achieve such a mapping using a latent representation where each latent component $z_i$ follows a Gaussian distribution that is independent of the others i.e., $p_{\phi} (\bol{z} | \bol{x}) = { \prod_{i=1}^L}  \mathcal{N}(\mu_i(\bol{x}), \sigma_i(\bol{x}))$, where $L$ is the dimensionality of the latent space. Under these assumptions, the encoder maps the inputs $\bol{x}$ to the vectors $\mu = {\mu_i, .., \mu_L}$ and $\sigma = {\sigma_i, .., \sigma_L}$. 

The decoder of the IVE consists of another neural network model, consisting of 3 fully-connected layers. A fully-connected layer is made of a number of neurons, such that every neuron in one layer is connected to every neuron in adjacent layers. Each neuron follows $y=h(\bol{w} \,\bol{x} + b)$, where $\bol{x}$ are the inputs, $y$ is the output, $h$ is the non-linear activation function and $\bol{w}, b$ are trainable parameters known as weights and biases.
Mathematically, the decoder consists of a model with parameters $\theta$ mapping a latent vector $\bol{z}$, sampled from $p_{\phi}(\bol{z} | \bol{x})$\footnote{In practice, the latent vector $\bol{z}$ is generated using the reparametrization trick i.e., $z_i = \mu_i + \sigma_i \epsilon_i$ where $\epsilon_i \sim \mathcal{N}(0,1)$, in order to preserve differentiability throughout the whole network.}, and a value of the query $\log(r)$ to a single predicted estimate for $\log[\rho_\mathrm{pred}(r)]$. Given many different realizations of $z \sim p_{\phi}(\bol{z} | \bol{x})$, one obtains a distribution of possible values for $\log[\rho_\mathrm{pred}(r)]$ as a function of $r$.

\subsection{The loss function}
Training the IVE requires solving an optimization problem. The parameters of the encoder and decoder, $\phi$ and $\theta$, are optimized to minimize a loss function which measures how close the predictions, $\bol{\rho}_\mathrm{predicted}$, are to their respective ground-truths, $\bol{\rho}_\mathrm{true}$, for the training data. The loss function is also designed to maximize the degree of independence of the latent variables; this is necessary in order to achieve our goal of a disentangled latent representation. These requirements can be obtained with the following loss function \cite{Higgins2017},
\begin{equation} 
\mathcal{L} = \mathcal{L}_\mathrm{pred}( \bol{\rho}_\mathrm{true}, \bol{\rho}_\mathrm{pred} )  + \beta \, \mathcal{D}_\mathrm{KL}[p_{\phi}(\bol{z} | \bol{x}); q(\bol{z})],
\label{eq:loss}
\end{equation}
where the first term measures the predictive accuracy of the model and the second is the Kullback-Leibler (KL) divergence \cite{Kullback1951} between the latent distribution returned by the encoder $p_{\phi}(\bol{z} | \bol{x})$  and a prior distribution over the latent variables $q(\bol{z})$. The parameter $\beta$ weights the KL divergence term with respect to the predictive term, and must be carefully optimized. We took the predictive term to be the mean squared error loss, 
\begin{equation}
\mathcal{L}_\mathrm{pred} = \frac{1}{N}\sum_{i=1}^{N} \left( \log_{10} \rho_{i,\mathrm{true}} - \log_{10} \rho_{i,\mathrm{pred}} \right)^2,
\label{eq:losspred}
\end{equation}
where $N$ is the training set size. Assuming a set of independent unit Gaussian distributions as the prior over the latent variables $q(\bol{z)}$, the KL divergence term takes the closed form,
\small
\begin{equation}
\mathcal{D}_\mathrm{KL}(\mathcal{N}(\mu_{\bol{z}}, \sigma_{\bol{z}}); \mathcal{N}(0, 1)) = -\frac{1}{2}\sum_{i=1}^L \left[ 1 + 2\log \sigma_i - \mu_i^2 - \sigma_i^2 \right].
\label{eq:KL}
\end{equation}
\normalsize
where $L$ is the dimensionality of the latent space.

The role of the KL term in the loss function is to promote independence between the latents \cite{Higgins2017}. This encourages the model to find a disentangled latent space, where independent factors of variation in the density profiles are captured by different, independent latents. Here, independence is intended in terms of both linearly and non-linearly uncorrelated variables. Hence, linear correlation measures such as the Pearson correlation coefficient are insufficient. We therefore evaluate the degree of disentanglement in terms of mutual information: the amount of shared information across the latents should be negligible if these describe independent factors of variation in the density profiles. 

Compressing the information in the input data into a disentangled low-dimensional latent representation can be thought of as a non-linear principal component analysis (PCA). PCA is a dimensionality-reduction technique to linear transform a set of correlated variables into linearly-uncorrelated components. The components describe linearly-uncorrelated factors of variability of the data set and dimensionality reduction is achieved by discarding components which describe negligible variability of the data. The encoder plays the role of a non-linear PCA: it performs a \textit{non-linear} transformation of the input data into \textit{disentangled} components. The degree of disentanglement of the latent space therefore crucially affects our ability to interpret the latent space in terms of independent factors of variability in the density profiles.

\subsection{Training the IVE model}
The training set consists of $\sim 6000$ halos per simulation for a total of three simulations, whereas the validation set consists of a random subset of 2000 halos from an independent simulation. The test set is given by all halos (also $\sim 6000$) from an independent simulation not used for training. We trained and validated the model in alternate radial bins of the profile, except the first and last radial bins that are used for both training and validating. This was done to ensure the model would not overfit to local fluctuations in the density profile but only learn global features of the dataset. The training set was sub-divided into batches, each made of $64$ halos. Batches were fed to the network one at a time, such that the model updates its parameters at every batch iteration.

Training was done using the \texttt{AMSGrad} optimizer \cite{Reddi2019}, a variant of the widely-used \texttt{Adam} optimizer \cite{Kingma2014}, with a learning rate of $5\times 10^{-3}$. We calibrated the parameter $\beta$ in the loss function of Eq.~\eqref{eq:loss} using cross-validation over a grid of 6 values in the range $[ 10^{-1}, 10^{-4}]$. Our aim was to select a value of $\beta$ that yielded both good predictive accuracy and a latent space with maximum disentanglement, defined as minimal mutual information between the different latents. Small values of $\beta$ prioritize the mean squared error term over the KL divergence term, yielding accurate predictions albeit an entangled latent space. Large values of $\beta$ instead yield a more disentangled latent space, at the expense of a degradation in the accuracy of the predictions. To select the best model, we tracked the values of the total loss function, the mean squared error and the KL divergence at every epoch during training. We started with an initial low value of $\beta$ so that the model first prioritized yielding the highest possible predictive accuracy; as the mean squared error stopped decreasing, we then gradually increased $\beta$ to promote a more disentangled latent space without significantly degrading the predictive performance. All models were run on 4 Nvidia A100 40GB GPUs. The total training time of the \IVEo{} model with a 3D latent space is $\sim 12$ hours on 4 Nvidia A100 GPUs. Once the model is trained, the density profile of a single halo can be obtained in $\sim 0.0076$ seconds from the 3D raw density field.

\section{Predicting the density profile of halos}
\label{sec:predictions}
\begin{figure}
\centering
\begin{subfigure}
\centering
\includegraphics[width=\columnwidth]{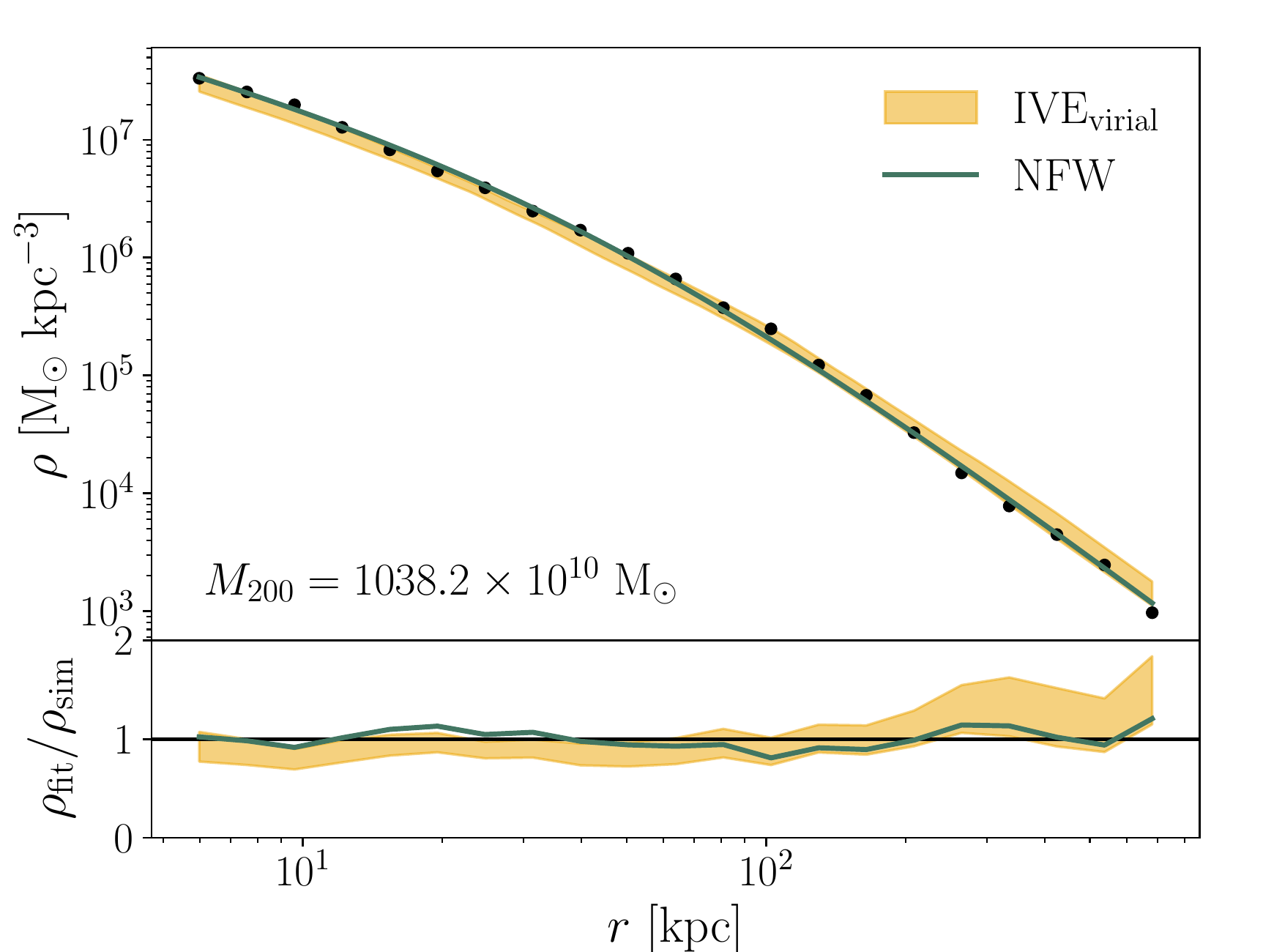}
\end{subfigure}
~
\begin{subfigure}
\centering
\includegraphics[width=\columnwidth]{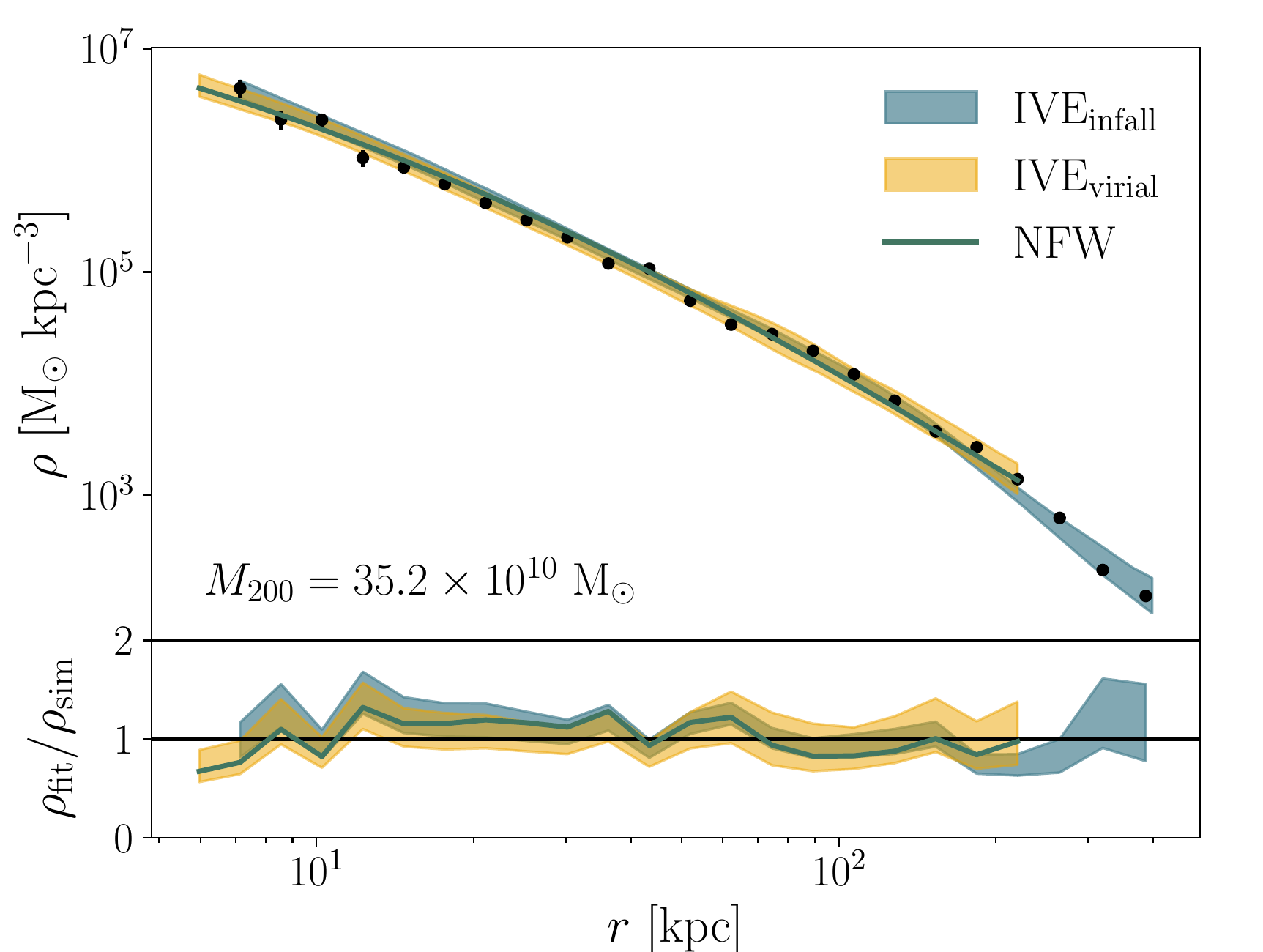}
\end{subfigure}
\begin{subfigure}
\centering
\includegraphics[width=\columnwidth]{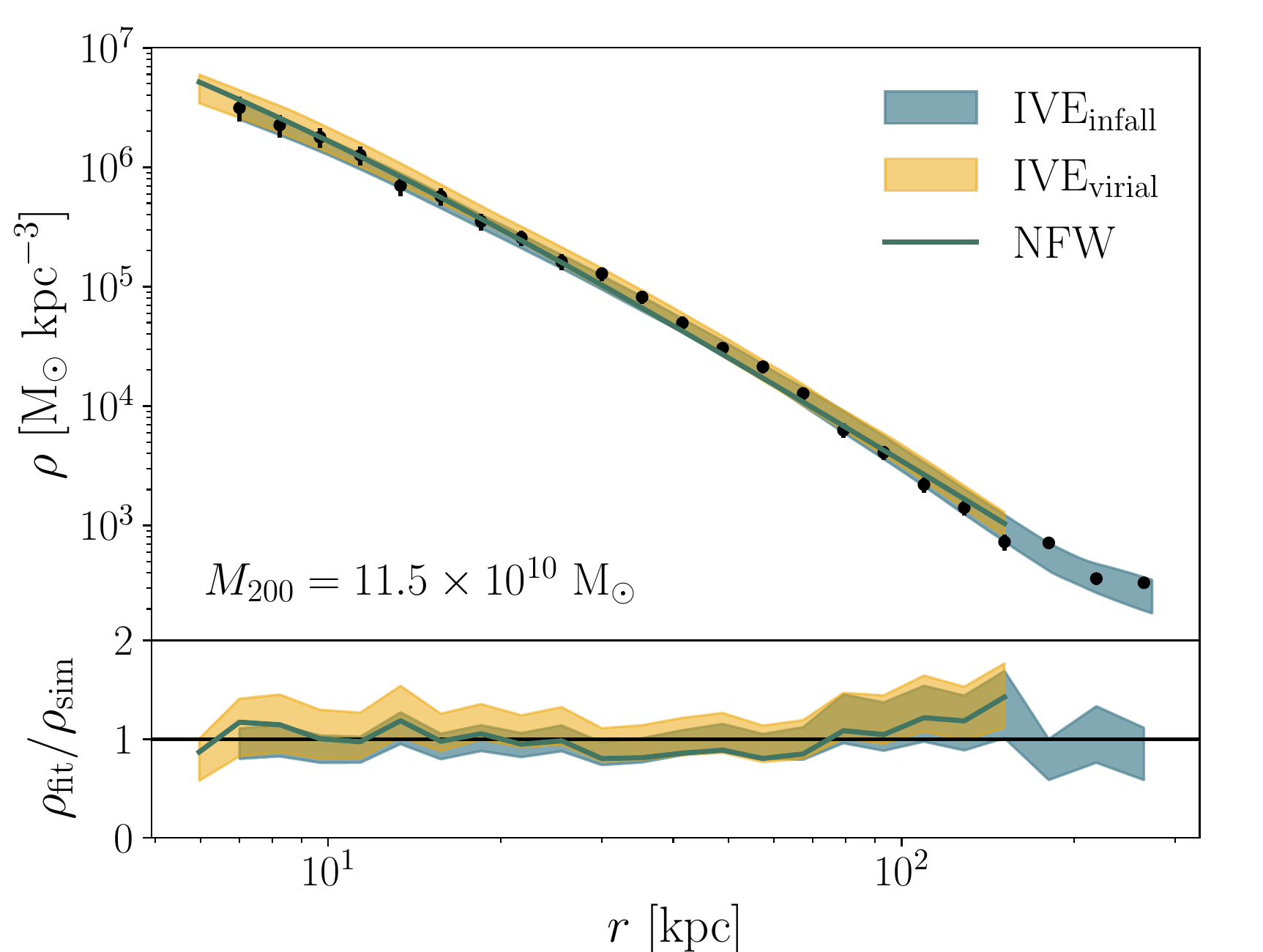}
\end{subfigure}
~
\caption{Three examples of fits to the density profiles of one high-mass (\textit{upper panel}), one mid-mass (\textit{middle panel}), and one low-mass (\textit{lower panel}) halo in the mass range considered. The IVE outputs a distribution of predicted density profiles given different random realizations of the latent representation. The NFW fit is also shown for comparison. For lower mass halos, we also show the predictions of the \IVEo{} model, trained to predict profiles out to large radii in the outskirts of halos.}
	\label{fig:haloexamples}
\end{figure}

Figure~\ref{fig:haloexamples} shows examples of fits to the density profiles of one high-mass (\textit{upper panel}), one mid-mass (\textit{middle panel}), and one low-mass (\textit{lower panel}) halo made by various models. The black points show the ground-truth density profiles from the simulations; the two coloured bands are the profiles predicted by the \IVEi{} and \IVEo{} models. The \IVEo{} model is only trained on lower-mass halos (see Sec.~\ref{sec:inputs}), thus we show predictions from that model only in the middle and lower panels. The IVE predictions are given by the range of predicted profiles given 100 random realizations drawn from the latent distributions. These three halo examples are representative of the diversity in the density profiles of our halo population: the profiles cover different dynamical ranges in radius, the upper-panel halo reaches a much larger inner density compared to the other two, and the outer profile of the lower-panel halo flattens out at large radii compared to the sharper decrease in the outskirts of the middle-panel halo.

We compare the IVE models to the most widely-used analytic fitting formulae for the density profiles of halos. The first model is the isotropic NFW profile \cite{Navarro1997}, a two-parameter functional form given by
\begin{equation}
\rho (r) = \frac{\rho_s}{r/r_s \left( 1 + r/r_s \right)^2},
\label{eq:nfw}
\end{equation}
where $r_s$ and $\rho_s$ are the scale radius, defined as the radius at which $\mathrm{d} \ln \rho/\mathrm{d} \ln r =-2$,  and the characteristic density, respectively. The second is the Einasto density profile \cite{Einasto1965}
\begin{equation}
\rho (r) = \rho_s \exp \left[ \frac{-2}{\alpha} \left( \left( \frac{r}{r_s}\right)^\alpha -1 \right) \right],
\end{equation}
where the additional parameter $\alpha$ is a shape parameter that regulates a smoother, more gradual transition between the two profile slopes compared to NFW. We consider the case where we fix $\alpha=0.16$, the best-fitting value found in previous literature for the mass range of halos we consider \cite{Gao2008}, and the case where we let $\alpha$ vary for every halo. We fitted NFW and Einasto formulae to the each halo's profile over the same radial bins used to train and validate the VAE model, by minimizing the expression:
\begin{equation}
\Psi^2 = \frac{1}{N_\mathrm{bin}} \sum_{i=1}^{N_\mathrm{bin}} \left[ \log_{10} \rho_\mathrm{sim, i} - \log_{10} \rho_\mathrm{fit, i}\right]^2,
\end{equation}
where $\log_{10} \rho_\mathrm{sim, i}$ and $\log_{10} \rho_\mathrm{fit, i}$ are the simulation's ground-truth data and the fitted density profile in radial bin $i$. This expression minimizes the rms deviation between the halos' binned $\rho(r)$ and the NFW profile, assigning equal weight to each bin. The solid lines in Fig.~\ref{fig:haloexamples} show the fits to the NFW profile for the three halo examples for comparison. The IVEs and NFW predictions both yield a good fit to the halo's density profiles. This demonstrates the ability of the IVE models to capture the diversity in the density profiles of our halo population, given a compact latent representation. 

\begin{figure}
	\centering
	\includegraphics[width=\columnwidth]{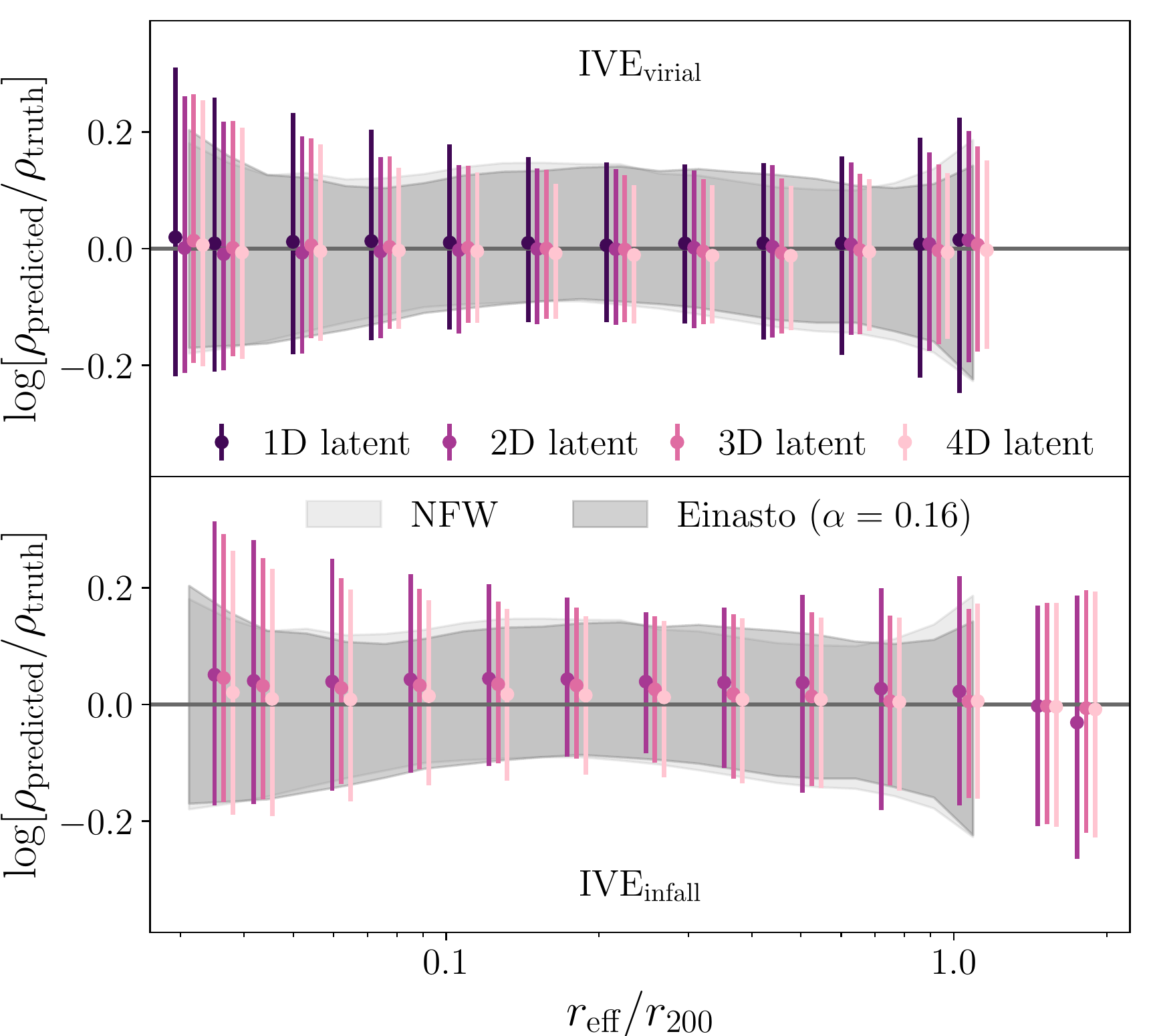}
	\caption{Mean and $90\%$ confidence interval of the residuals, $\log[\rho_\mathrm{predicted}/\rho_\mathrm{truth}]$, for the \IVEi{} and \IVEo{} models, trained using different choices of latent dimensionality. The light and dark grey bands shows the NFW and Einasto residuals, respectively. \textit{Upper panel}: Models trained up to the halo boundary require a 2D latent space to yield accurate predictions that are consistent with analytic models. \textit{Lower panel:} Modeling the density profiles out to large radii requires a third parameter to model the component of infalling material in the outskirts of halos.}
	\label{fig:CI_different_dim}
\end{figure} 

We compare the predictions of the IVE models, trained with different latent dimensionalities, and those of the analytic models in Fig.~\ref{fig:CI_different_dim}. We show the mean and $90\%$ confidence interval of the residuals $\log[\rho_\mathrm{predicted}/\rho_\mathrm{truth}]$, in every radial bin of the profile used for testing. Each radial bin corresponds to a different value of $r$ for different halos; we therefore define $r_\mathrm{eff}$ to be the median of the distribution of radius values within each bin. We use this quantity throughout the paper. The different colours show the residuals of IVE models with different choices of latent space dimensionality; the grey bands shows the residuals of the NFW model and the Einasto model with fixed $\alpha$. The top panel shows the \IVEi{} case. We find that increasing the latent dimensionality to more than two parameters yields no significant improvement in the predictions. On the other hand, an IVE with 1-dimensional latent space yields larger errorbars than all other models, especially in the innermost and outermost radial bins. This implies that a two-parameter model is sufficient to capture the diversity of the density profiles of individual halos within the virial radius. Moreover, since the performance of the 2D (or more) IVE is consistent with that of existing well-known models such as the NFW and Einasto profiles, we conclude that our models contain sufficient predictive accuracy to yield meaningful interpretations of their latent representations.

The bottom panel shows the \IVEo{} case. Here, a 2D latent space model yields larger residuals in both the inner profile and the region close to the virial radius. As we increase the latent space to 3 parameters, the model reduces the scatter close to the virial boundary of the halo, and reduces the bias in the predictions of the inner profile. Further increasing the latent space dimensionality to 4 parameters does not yield significant improvement. This demonstrates that one additional parameter is required to model halo density profiles up to the halo outskirts, compared to the number of parameters required to describe profiles within \rvir{}\footnote{We note that \IVEo{} mildly increases the size of the residuals in the innermost region of the profiles, when compared to the NFW and Einasto cases.}.

\section{Interpreting the latent representation}
\label{sec:latents}
\begin{figure*}
	\centering
	\includegraphics[width=\textwidth]{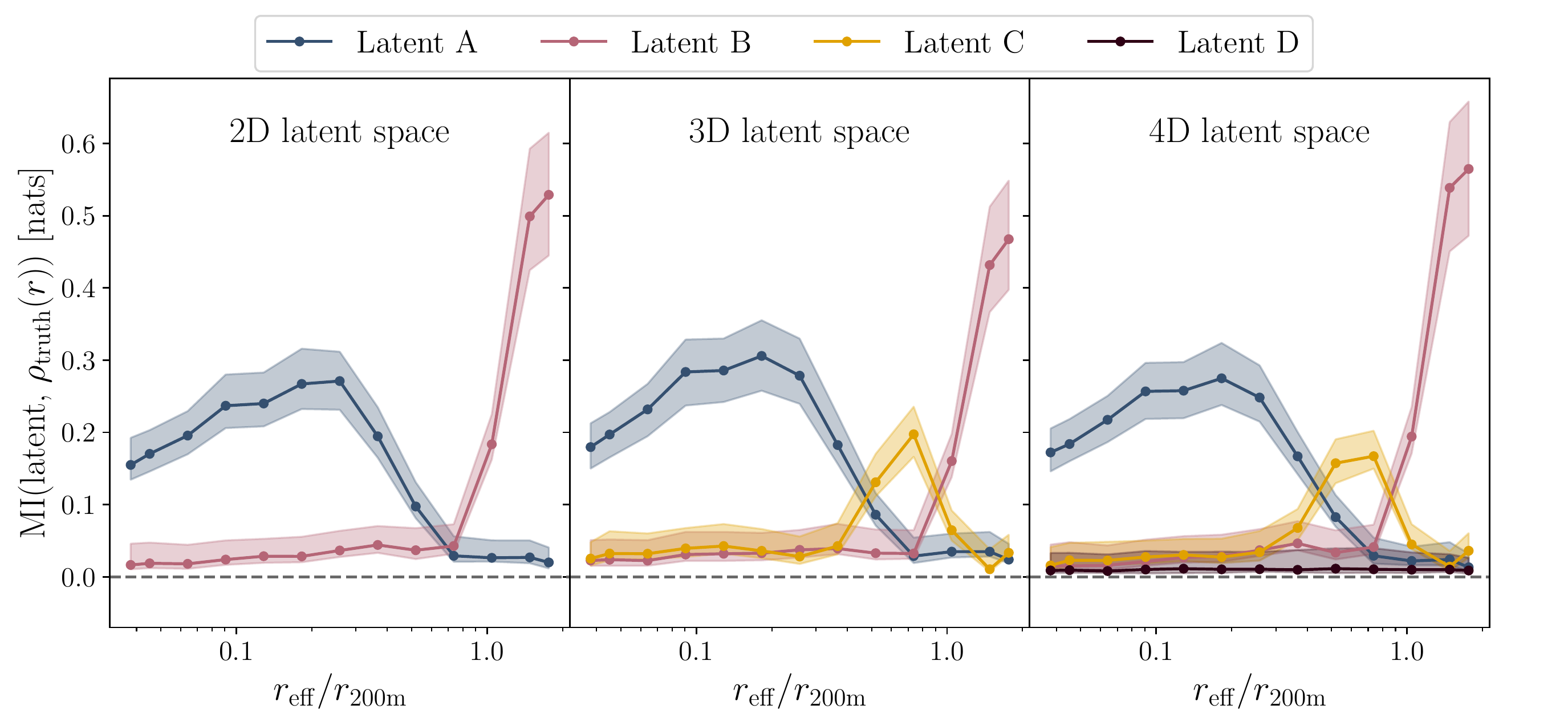}
	\caption{Mutual information between each latent variable and the ground-truth $\log[\mathbf{\rho}_i(r)/\bar{\rho}_\mathrm{m}]$ in every $i$-th radial bin. The three panels show the results for the \IVEo{} model with latent dimensionality 2, 3 and 4. The solid lines show the mutual information when adopting a bandwidth of $0.2$; the bands show the scatter in the mutual information estimate when adopting bandwidths of 0.1 (upper band limit) and 0.3 (lower band limit). These values of bandwidths cover sufficient range to undersmooth and oversmooth the distributions, thus demonstrating that our results are insensitive to the specific bandwidth choice.}
	\label{fig:MI_different_latent}
\end{figure*} 

The information contained within the input sub-box of every halo in the test set is compressed into $L$ latent Gaussian distributions by the encoder. The decoder draws samples from these latent distributions to produce density profile predictions for the halos. The latent distributions therefore contain the necessary information to predict the density profile of a given halo. We quantified the information contained within the IVE latent variables about the ground-truth density profiles by estimating the mutual information between the two quantities, as follows.

\subsection{Estimating mutual information}
Our goal is to evaluate the mutual information between each latent and the ground-truth density as a function of $r$. To do so, we required two quantities: (i) the distribution of possible latent realizations over all halos, (ii) the distribution of ground-truth density values $\rho(r)$ as a function of $r$.

We estimated quantity (i) as follows. Each halo's latent representation is defined by $\mu_{j,\alpha}$ and $\sigma^2_{j, \alpha}$, the mean and variance of the $j$-th latent Gaussian distribution for halo $\alpha$. We drew one sample from each halo's latent distribution, $z_{j,\alpha} \sim \mathcal{N}(\mu_{j,\alpha}, \sigma_{j, \alpha})$, which in turn yielded a set of latent variables $\bol{z}_j = \{ z_{j,\alpha} \}_{\alpha=1}^M$ for all $M$ halos. We turned the discrete distribution constructed from $\bol{z}_j$ into a continuous probability density function using a kernel density estimation (KDE) method \cite{rosenblatt1956}. A KDE is a non-parametric approach to estimate the probability density distribution from a discrete set of samples. Each data point is replaced with a kernel of a set width and the density estimator is given by the sum over all kernels. For the case of the $M$ discrete values of latent samples $z_j$, its kernel density estimate is given by
\begin{equation}
p(z_j) = \frac{1}{M} \sum_{\alpha=1}^{M} K\left( \frac{z_j - z_{j, \alpha}}{b} \right),
\label{eqref:kde}
\end{equation}
where $K$ is the kernel, which we take to be a Gaussian of the form $K(x) \propto \exp(-x^2/2)$, and $b$ is a free parameter known as the bandwidth, which determines the width of the kernel. The bandwidth is a free parameter which must be tuned to the distribution at hand: if too small, the density estimate will be undersmoothed and noisy; if too large, the density estimate will be oversmoothed and may wash out important features of the underlying structure. To estimate quantity (ii), we took the set of halo ground-truth densities $\bol{t}_i = \{ \log[\rho_{i, \alpha}(r)] \}_{\alpha=1}^M$ in radial bin $i$ for all $M$ halos. As before, we turned the discrete sampling distribution constructed from $\bol{t}_i$ into a continuous probability density function, $p(t_i)$, using a KDE as in Eq.~\eqref{eqref:kde}.

Finally, the mutual information (MI) between latent $j$ and the ground-truth density in radial bin $i$ is given by
\begin{equation}
\mathrm{MI}(t_i, z_j) = \int_{t_i} \int_{z_j} p(t_i, z_j) \log \left[ \frac{p(t_i, z_j)}{p(t_i) p(z_j)} \right] dt_i dz_j,
\end{equation}
where $p(t_i, z_j)$ is the joint probability density function between $t_i$ and $z_j$. This was computed using a 2D KDE method, similar to the 1D probability density functions $p(t_i)$ and $p(z_j)$. The limits of integration are the minimum and maximum values of $t_i$ and $z_j$, respectively.

\subsection{Mutual information between latents and halo profiles}
Figure \ref{fig:MI_different_latent} shows the mutual information between the latent variables and the ground-truth density in every radial bin, for the \IVEo{} model with a 3D latent space. Each curve represents the mutual information between one latent $z_j$ and the distribution of $\log[\mathbf{\rho}_i(r)]$ in each radial bin $i$. The solid lines show the mutual information when adopting a bandwidth of $0.2$ for the KDE fit; the bands show the scatter in the mutual information estimate when adopting bandwidths of 0.1 (upper band limit) and 0.3 (lower band limit). These choice of bandwidths were made to cover a large bandwidth range, from small values that undersmooth the distributions to large values where the distributions are oversmoothed. These scales are set by the dynamic range of values covered by the distributions: a bandwidth of 0.1 yields a multi-modal distribution function with large numbers of peaks, whereas one of 0.3 yields a single-peaked distribution function. This demonstrates that our results are insensitive to the specific choice of bandwidth in the KDE in Eq.~\eqref{eqref:kde}. Each panel shows the results for the \IVEo{} model with different latent dimensionality. 

Latent A encodes similar information about the density profiles in all models: it accounts for the largest component of variability in the profiles within the virial radius. In particular, its mutual information with $\rho(r)$ peaks at scales $r\sim 0.3~$\rvir. Latent B contains information about the outer profile beyond \rvir. Two components are not sufficient to correctly model the full diversity of halos, as seen in Fig.~\ref{fig:CI_different_dim}. A third latent (latent C) must be included to account for additional variability in  the density profiles on scales approaching \rvir. This is consistent with the finding in the bottom panel of Fig.~\ref{fig:CI_different_dim}, showing an improvement in the residuals of the predictions on similar scales when going from the 2D to the 3D latent space models. On the other hand, the addition of a fourth latent variable has very little effect on the predictivity of the model; this result is also reflected in the fact that the fourth latent has negligible mutual information with the ground-truth density field at all radii. These results also explain why a two-parameter model is sufficient to accurately model the profiles within \rvir (upper panel of Fig.~\ref{fig:CI_different_dim}); the third component is only needed to describe the halo profile outskirts. We further confirmed this result by estimating the mutual information between the \IVEi{} latents and the ground truth density profiles; using a three-dimensional latent space, we found two latents encoding the same information as latent A and latent C in Fig.~\ref{fig:MI_different_latent} and a third latent encoding no significant information about the profiles.

We also measured the mutual information between the latents themselves, to test whether any information about the halo profiles is shared amongst the latents. We find that the mutual information between latents is $\mathcal{O}(10^{-2})$ nats, confirming that the IVE has found a disentangled latent representation of halo density profiles, such that each latent captures different, independent factors of variation in the profiles.

\begin{figure*}
	\centering
	\includegraphics[width=\textwidth]{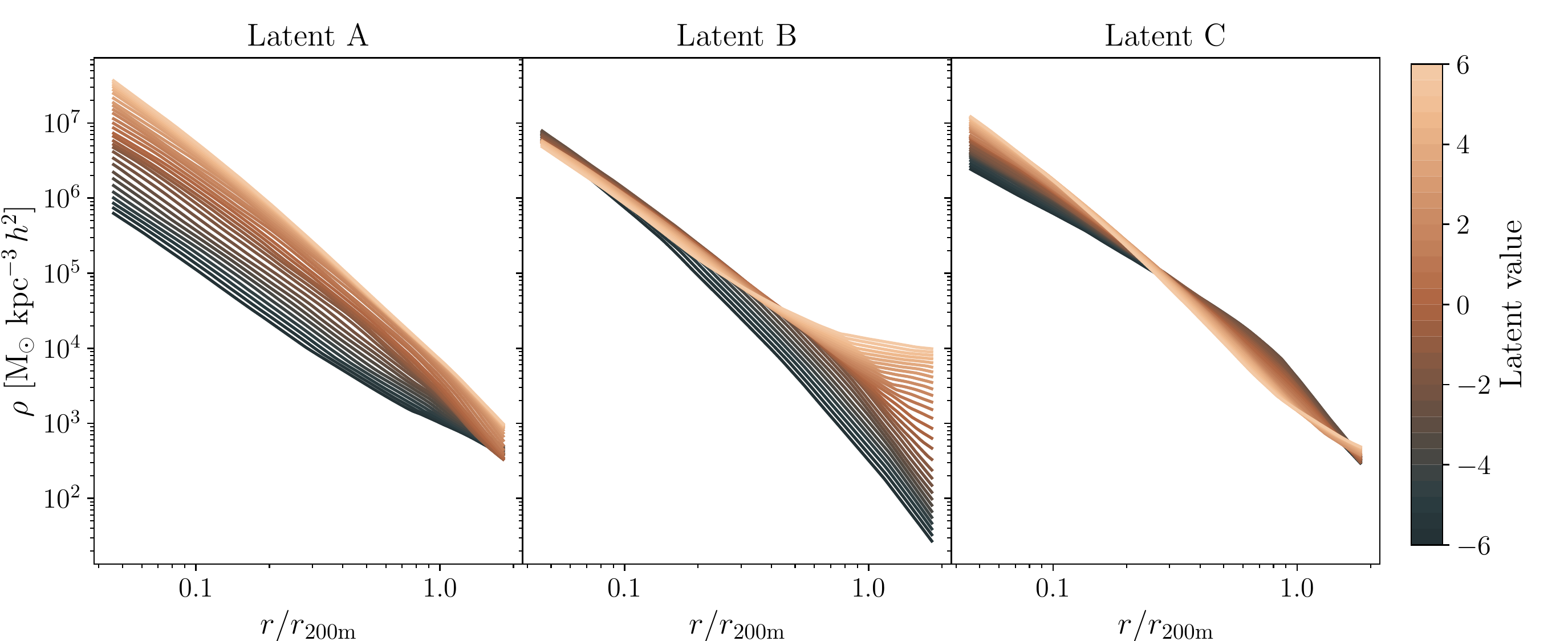}
	\caption{Variations in the predicted density profile of a given halo when systematically varying the value of one latent, while keeping the others fixed. Each panel from left to right varies latent A, B or C, respectively, as defined in Fig.~\ref{fig:MI_different_latent}. The first latent describes the normalization of the profile; the second the shape of the outer profile component; the third the steepness, or shape, of the profile within \rvir{}. These independent aspects of the halos' density profiles were discovered automatically by the IVE during training.}
	\label{fig:vary_latents}
\end{figure*}

We further investigated the information content of the latent representation discovered by the 3D latent space \IVEo{} for individual examples. Given the latent distributions returned by the encoder for any given halo, we systematically varied the value of one latent, while keeping the others fixed to the mean of their respective Gaussian distributions. This allowed us to directly probe how the predicted density profile varies as its latents vary one at a time. Fig.~\ref{fig:vary_latents} shows the variation in the predicted density profile of a halo as we systematically change the values of latent A, B or C independently, each time keeping the other two fixed. Latents A, B, and C describe the normalization of the whole profile, the shape of the outer profile, and the shape of the profile out to \rvir{}, respectively. The shape of the outer profile at the transition between orbiting and infalling material into the halo is determined by the splashback radius i.e., the location where particles reach the apocenter of their first orbit \cite{Diemer2014, Adhikari2014, More2015, Shi2016}. Therefore, the variability in the outer profile captured by latent C can be thought as variability in the location of the splashback radius of the halos. This implies that the \IVEo{} was able to discover the splashback feature in dark matter halos, without prior knowledge of the halos' dynamical history.

Varying one latent at a time (Fig.~\ref{fig:vary_latents}) provides us with a different, but related, perspective on the information content of the latent representation compared to the mutual information measure (Fig.~\ref{fig:MI_different_latent}). The former shows how the predicted profiles depend on any given latent, conditioned on fixed values of the other latents; the mutual information reveals a more global dependency between latents and ground-truths, sensitive to variations in the profiles from all factors simultaneously. For example, the mutual information of latent A with $\rho(r)$ peaks at $r\sim 0.3~$\rvir{} and decreases rapidly as we move towards the outskirts (Fig.~\ref{fig:MI_different_latent}), despite capturing information about the normalization of the profile which affects the profile at all radii equally (Fig.~\ref{fig:vary_latents}). This indicates that the variability in the inner profile is dominated by the variability in the normalization, whereas the outskirts of the profiles are dominated by other factors --- e.g., infalling material captured by latent B. The two techniques together — quantifying the mutual information and scanning through the latents one at a time — provide complementary ways to interpret the latent space.

\subsection{Comparing latent and analytic representations}
\label{sec:analytic}
\begin{figure}
	\centering
	\includegraphics[width=\columnwidth]{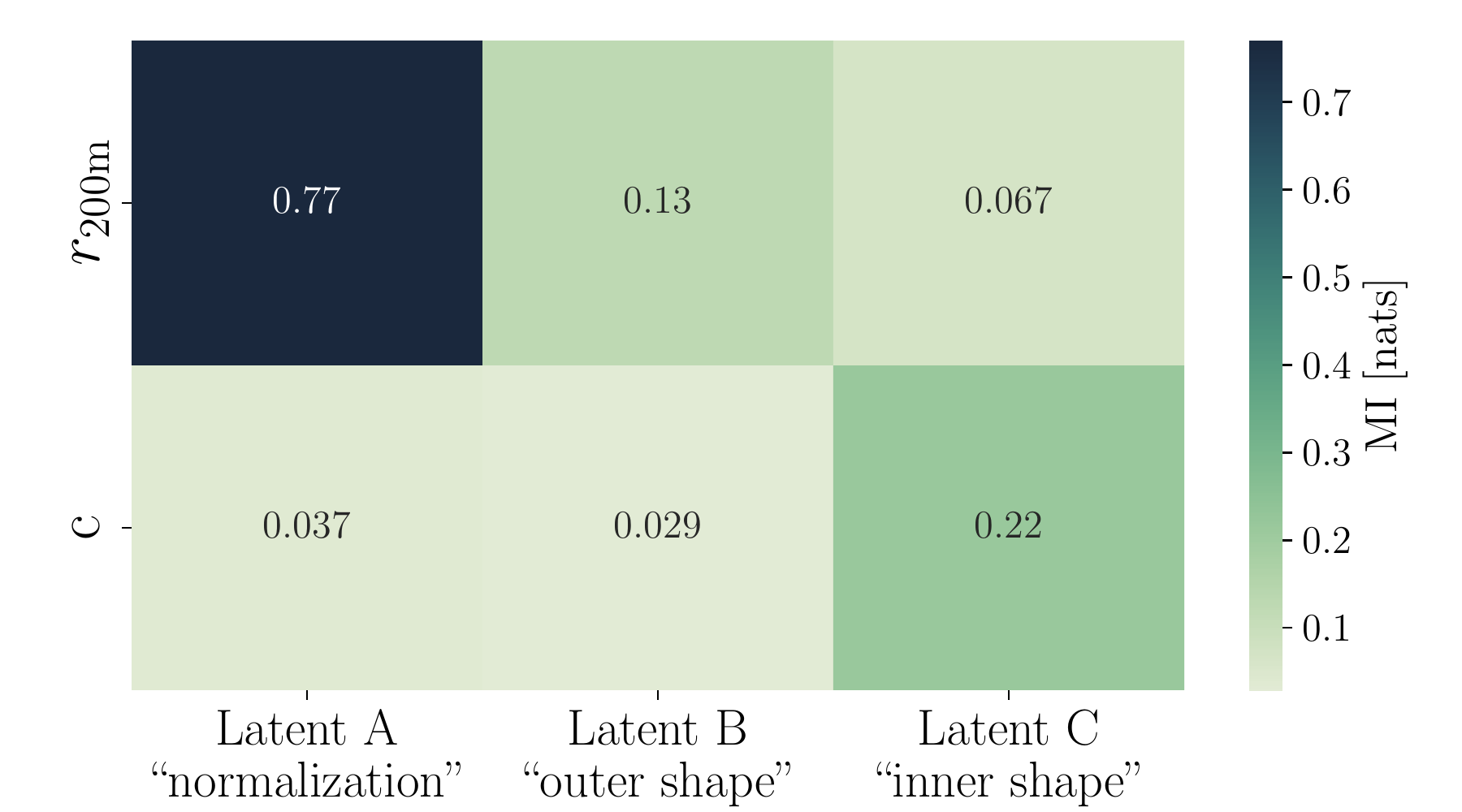}
	\caption{Mutual information between the latent parameters learnt by \IVEo{} and the two parameters adopted by the NFW model. Latent A contains a large amount of shared information with \rvir{}, since both parameters describe the normalization of the profile. Latent C shares some information with concentration, as these both describe the profile shape. Most of the information captured by latent B is not present in the NFW parameters. However, there is some shared correlation with \rvir{}, which is expected since the outer profile should also depend on the size of the halo.}
	\label{fig:nfw_vs_vae_latents}
\end{figure}
We next compare the information stored within the latent representation to that stored in the parameters of the NFW profile, by computing their mutual information as shown in Fig.~\ref{fig:nfw_vs_vae_latents}. The NFW profile, written as a function of the scale radius $r_{s}$ and the characteristic radius $\rho_{s}$ as in Eq.~\eqref{eq:nfw}, can be re-expressed in terms of the virial radius \rvir{} and the concentration $c=$ \rvir{}$/r_s$. Latent A contains a large mutual information with \rvir{}: indeed, both parameters are responsible for the normalization of the profile as corroborated in Fig.~\ref{fig:MI_different_latent} \& \ref{fig:vary_latents}. Latent C shares information with concentration: both of these describe the shape of the profile. However, the amount of shared information between concentration and latent C is not as high as that between latent A and \rvir{}. This suggests that latent C and concentration affect the shape of the profiles differently. The outer profile is not well-modeled by the NFW profile and therefore most of the information captured by latent B is not present in the NFW parameters; however, there is a weak amount of shared correlation between latent B and \rvir{} since we expect the outer profile to also depend on the size of the halo.

\section{Conclusions}
\label{sec:conclusions}
We have presented an interpretable encoder-decoder framework, capable of predicting the spherically-averaged density profiles of halos given the raw 3D density field containing each halo. Our goal is to disentangle the independent degrees of freedom in cosmological dark matter halo density profiles. Our model consists of an encoder mapping the density field containing each halo to a low-dimensional latent representation, followed by a decoder mapping the latent representation and an input value of $r$ to the spherically-averaged density $\rho(r)$. This architecture was specifically designed to extract the building blocks of dark matter halo density profiles from the neural network: all the information used by the model to predict density profiles is stored in the representation, and the size of this representation is small compared to the total number of parameters in the network. We interpret the latent representation by quantifying the mutual information between each latent dimension and the halos' density profiles.

We find that a two-dimensional representation is sufficient to accurately model the density profiles up to the virial radius; however, a three-dimensional representation is required to describe the outer profiles beyond the virial radius. We show that the machine-learning model, using the full 3D density field containing the halo as input, discovers similar quantities to those used by the well-known NFW profile. One latent component describes the overall normalization of the profiles, and the second the steepness of the profile within the virial radius. The third latent contains information about the infalling material in the outskirts of dark matter halos. Therefore, the machine-learning model discovers the splashback boundary of halos without prior knowledge about the halos' dynamical histories.

The IVE provides us with an alternative compact parametrization of halo density profiles. This can be particularly useful for studies of the transition between orbiting and infalling material into the halo, including the location of the splashback radius. The outer profile is difficult to model analytically due to its intrinsically dynamical nature \cite{Diemer2014, Adhikari2014, Diemer2021}. In addition to providing an alternative `fitting' function for halo density profiles, our result that three parameters (and not more) are needed to describe profiles beyond the virial radius can also inform future theoretical studies on the origin of density profiles from first principles.

Our approach presents a generalization over existing analytic fitting formulae, such as the NFW and Einasto profiles, which are typically fitted to preprocessed information about the halos' structure i.e., the spherically-averaged density profiles themselves. Thus, our machine-learning framework is not limited to spherically-averaged densities, but straightforwardly generalizes to other halo observables that target for example halo triaxality or substructures. Our IVE framework can also be used to study the impact of baryonic physics on the dark matter halos' density profiles by training the model using hydrodynamical simulations. Moreover, the machine-learning model is trained to predict the density profile of any individual halo found in a $\Lambda$CDM simulated universe; on the other hand, existing fitting formulae are typically fitted to the stacked profiles of populations of relaxed halos.

This work makes progress toward designing interpretable machine-learning frameworks for extracting new knowledge about the underlying physics of cosmological structure formation. In particular, the IVE architecture allows one to automatically generate a compact model, captured within a disentangled latent representation, that retains all the information in the input data needed to predict a given property of interest. Mutual information then provides a principled quantitative interpretability measure to relate this compact representation to physical factors underlying the mapping. This framework is broadly applicable to other problems in cosmological structure formation, such as modelling the halo mass function, the void size function or the void density profile.

\section*{Author Contributions}
\textbf{L.L.-S}: led the project; project conceptualization; methodology; software; obtained, validated, and interpreted results; writing - original draft, editing, final; visualisation.
\textbf{H.V.P.}: project conceptualization; methodology; investigation, validation \& interpretation; writing - editing; funding acquisition.
\textbf{A.P.}: project conceptualization; methodology; investigation, validation \& interpretation; writing - editing; funding acquisition.
\textbf{B.N.}: methodology; investigation, validation \& interpretation; writing - editing.
\textbf{J.T.}: writing – editing.
\textbf{D.P.}: writing – editing.

\begin{acknowledgments}
LLS thanks Susmita Adhikari, Sten Delos, Eiichiro Komatsu, Martin Rey, Jie Wang and Simon White for useful discussions. LLS further thanks Volker Springel and Ruediger Pakmor for guidance with GADGET-4. HVP thanks Daniel Mortlock for useful discussions about information-theoretic measures. HVP was partly supported by the research project grant “Fundamental Physics from Cosmological Surveys” funded by the Swedish Research Council (VR) under Dnr 2017-04212. The work of HVP was also supported by the G\"{o}ran Gustafsson Foundation for Research in Natural Sciences and Medicine. This project has received funding from the European Research Council (ERC) under the European Union’s Horizon 2020 research and innovation programme (grant agreement nos. 818085 GMGalaxies and 101018897 CosmicExplorer). AP was additionally supported by the Royal Society. DP was supported by the UCL Provost’s Strategic Development Fund. HVP acknowledges the hospitality of the Aspen Center for Physics, which is supported by National Science Foundation grant PHY-1607611. The participation of HVP at the Aspen Center for Physics was supported by the Simons Foundation. This manuscript has been authored by Fermi Research Alliance, LLC under Contract No. DE-AC02-07CH11359 with the U.S. Department of Energy, Office of Science, Office of High Energy Physics. This work was partially enabled by funding from the UCL Cosmoparticle Initiative. JT’s work was supported by the EPRSRC grants EP/T10001569/1 via Alan Turing Institute, and EP/V001310/1.
\end{acknowledgments}

\bibliography{paper}

\end{document}